# Redundant Logic Insertion and Fault Tolerance Improvement in Combinational Circuits

P. Balasubramanian
School of Electrical and Electronic Engineering
Nanyang Technological University
Singapore
e-mail: balasubramanian@ntu.edu.sg

R. T. Naayagi
School of Electrical and Electronic Engineering
Newcastle University International Singapore
Singapore
e-mail: naayagi.ramasamy@ncl.ac.uk

*Abstract*—This paper presents a novel method to identify and insert redundant logic into a combinational circuit to improve its fault tolerance without having to replicate the entire circuit as is the case with conventional redundancy techniques. In this context, it is discussed how to estimate the fault masking capability of a combinational circuit using the truth-cum-fault enumeration table, and then it is shown how to identify the logic that can introduced to add redundancy into the original circuit without affecting its native functionality and with the aim of improving its fault tolerance though this would involve some trade-off in the design metrics. However, care should be taken while introducing redundant logic since redundant logic insertion may give rise to new internal nodes and faults on those may impact the fault tolerance of the resulting circuit. The combinational circuit that is considered and its redundant counterparts are all implemented in semi-custom design style using a 32/28nm CMOS digital cell library and their respective design metrics and fault tolerances are compared.

## I. INTRODUCTION

The ITRS design report [1] has labelled reliability (which includes fault tolerance) as an important design challenge for nanoelectronics due to several complex technological issues associated with the continuous shrinkage of semiconductor device dimensions [2]. To improve the reliability and fault tolerance of electronic circuits and systems, redundancy has been widely resorted to in many mission-critical and safety-critical applications such as space, aerospace, nuclear, defense, banking and financial, any many other industrial applications such as electrical power transmission and distribution systems, data centers, cloud servers etc. [3].

Conventionally, redundancy involves duplication of the entire system or a portion of the system (called as sub-system or circuit). The classic N-modular redundancy (NMR) [3], [4] scheme involves deployment of (N – 1) copies of a circuit or system, and the outputs of the original circuit or system and its identical copies are combined using a voting element [4] which performs majority voting to determine the correct primary outputs. For example, in triple modular redundancy [5], [6] abbreviated as TMR, which forms a subset of the NMR [4], three copies of a circuit or system are used and a majority voting [7] is performed to determine the correct output(s) of the triplicated circuit or system. The TMR can successfully mask the faulty or failure state of any arbitrary circuit or system and hence its fault tolerance is said to be unity. Reference [8] proposed a novel current-based voting strategy for the NMR method which did away with the majority voter but [8] also entails the replication of the entire circuit or system according to the NMR approach, and so it is expensive. To avoid duplicating an entire circuit or system twice according to the TMR, approximate TMR has been proposed as an alternative [9]–[12] which leads to a partial TMR that facilitates reduced design cost and design metrics while involving an acceptable compromise on the TMR accuracy. However, a circuit or system considered for the approximate TMR may require approximately thrice the area of the original circuit or system. Nonetheless, approximate TMR is still being researched and its efficacy is yet to be demonstrated on real-time mission-critical and safety-critical circuits or systems.

Generally for enhanced fault tolerance, more copies of a circuit or system have to be added resulting in adoption of higher versions of the NMR [13]. Though higher versions of the NMR lead to enhanced reliability and fault tolerance, they also tend to exacerbate the design weight and cost and also the design metrics. To minimize these, recently [14], the distributed minority and majority voting based redundancy (DMMR) scheme was proposed. It was shown that DMMR could significantly reduce the design weight and cost and also the design metrics [15], [16] due to lesser duplication of the original circuit or system. Nevertheless, the NMR or DMMR scheme is still expensive for generic applications and may be suitable for only a selective deployment in mission-critical and safety-critical applications [17]. It is worth noting here that redundancy with respect to the NMR or the DMMR implies the entire duplication of the original circuit or system and not just the introduction of some extra logic i.e. redundant logic which does not affect the native functionality and only helps to increase the fault tolerance.

Other logic redundancy techniques which were proposed to improve the fault tolerance within certain bounds include interwoven redundant logic [18], quadded logic [19], [20], quadded transistors [21], quadded logic-cum-quadded transistors [22], but these tend to involve up to four times the logic density of a normal circuit implementation which may be expensive for arbitrary digital circuit implementations.

While reliability and fault tolerance of electronic circuits and systems have been the main forte of redundancy, there exists other benefits of introducing redundancy into a circuit or system. For example, logic redundancy has been used to achieve area and/or timing optimization in synchronous and asynchronous digital circuit designs [23], [24], and used to improve the yield of digital integrated circuit designs [25]. Reference [26] discusses a novel logic redundancy method called gate level information flow tracking, which consumes three times the area of the original circuit. Reference [27] also presented a new logic redundancy method called turtle logic, which primarily addresses the design of digital circuits to cope with noisy scenarios for operation under ultralow supply voltages (i.e. the subthreshold regime). Turtle logic, similar to that of quadded logic, basically addresses redundancy at the transistor level but the main drawback of turtle logic is that it is exorbitantly expensive and may not be preferable for generic digital circuit designs. For example, an inverter designed using turtle logic requires 20 transistors as opposed to just 2 transistors in the case of static CMOS logic. Moreover, the turtle logic necessitates the provision of complementary outputs for each gate functionality, which does not conform to commercial digital cell libraries, and therefore the turtle logic would require the development of a dedicated digital logic synthesis flow and hence may not be able to harness the power and maturity of electronic design automation implicit in commercial logic synthesis tools.

Given the above discussions, it may be rather clear that logic redundancy has primarily meant the duplication of an entire circuit (or sub-system or system), and not just the addition of some extra logic to intrinsically enhance the fault tolerance of the original circuit, which may form a part of a sub-system or a system. In this context, this paper presents a preliminary insight into a new logic redundancy scheme that deals with the systematic identification and inclusion of redundant logic into a combinational circuit, which would pave the way for improving its inherent fault masking capability without changing its native functionality.

The rest of this paper is organized into four sections. Section 2 considers an example combinational circuit and describes how to estimate its fault masking capability on the basis of the fault masking ratio by using the truth-cum-fault enumeration table proposed in [28]. In Section 3, it is discussed how to identify the redundant logic that may be inserted into the original circuit to improve its intrinsic fault tolerance without affecting its native functionality. Also, various fault tolerance enhanced implementations of the original circuit are presented. Section 4 provides the fault masking ratios of the different implementations and their respective design metrics. Lastly, the conclusions and scope for further work are suggested in Section 5.

## II. COMBINATIONAL CIRCUIT – CASE STUDY

Let us consider a 4-input combinational circuit which has 1 output, and this will be used as a running example in this work. The truth table of the combinational circuit is shown in Table I. The circuit property is that whenever two or more of its inputs are 1, its output is 1. The inputs are represented by A, B, C and D, and the output is represented by V.

TABLE I. TRUTH TABLE OF THE EXAMPLE COMBINATIONAL CIRCUIT

| Primary Inputs | | | | Primary Output |
|---|---|---|---|---|
| A | B | C | D | V |
| 0 | 0 | 0 | 0 | 0 |
| 0 | 0 | 0 | 1 | 0 |
| 0 | 0 | 1 | 0 | 0 |
| 0 | 0 | 1 | 1 | 1 |
| 0 | 1 | 0 | 0 | 0 |
| 0 | 1 | 0 | 1 | 1 |
| 0 | 1 | 1 | 0 | 1 |
| 0 | 1 | 1 | 1 | 1 |
| 1 | 0 | 0 | 0 | 0 |
| 1 | 0 | 0 | 1 | 1 |
| 1 | 0 | 1 | 0 | 1 |
| 1 | 0 | 1 | 1 | 1 |
| 1 | 1 | 0 | 0 | 1 |
| 1 | 1 | 0 | 1 | 1 |
| 1 | 1 | 1 | 0 | 1 |
| 1 | 1 | 1 | 1 | 1 |

After logic minimization, the reduced sum-of-products (SOP) form of V is given as,

$$V = AB + AC + AD + BC + BD + CD \quad (1)$$

After logic factorization [29], (1) reduces to (2). Equation (2) resembles a mixed logic synthesis form [30] that contains both sum of product terms and product of sum terms. The logic circuit corresponding to (2), synthesized using the elements of the 32/28nm CMOS cell library [31] is shown in Figure 1a. Alternatively, the product of sums (POS) form for V could be obtained and subsequent to logic factorization, the compact POS form of V is expressed by (3). The circuit synthesized corresponding to (3) is shown in Figure 1b. Note that $V_{SOP}$ and $V_{POS}$ are logically equivalent to V.

$$V_{SOP} = AB + CD + (A + B)(C + D) \quad (2)$$

$$V_{POS} = (AB + C + D)(A + B + CD) \quad (3)$$

Figures 1a and 1b consist of simple and complex gates of the cell library [31]. Figure 1a contains two complex gates viz. OA22 and AO221. With J being the output of the OA22 gate in Figure 1a, which is highlighted by the cross mark, the OA22 gate implements $J = (A + B)(C + D)$, and the AO221 gate implements $V = AB + CD + J$. Figure 1b contains two complex gates (AO221) and one simple gate (2-input AND). The AO221 gates implement $N1 = AB + C + D$ and $N2 = CD + A + B$, with N1, N2 being the intermediate outputs which are highlighted by the cross marks. Although AO211 gates are sufficient to implement N1 and N2, the cell library considered [31] does not have the AO211 gate. It was noted in [32] that the POS form of a logic function may lead to a compact circuit than the SOP form in certain cases and may facilitate optimization of the design metrics, but for our example circuit the SOP form appears to yield an optimized solution. This is because the factorized SOP form (2) after synthesis requires just two gates, whereas the factorized POS form (3) after synthesis requires an extra gate in comparison.

However, a direct comparison of (2) and (3) in terms of the number of gates may not lead to a definitive conclusion, and hence an estimation of the design metrics viz. power, delay, and area is necessary, and this is discussed in Section 4.

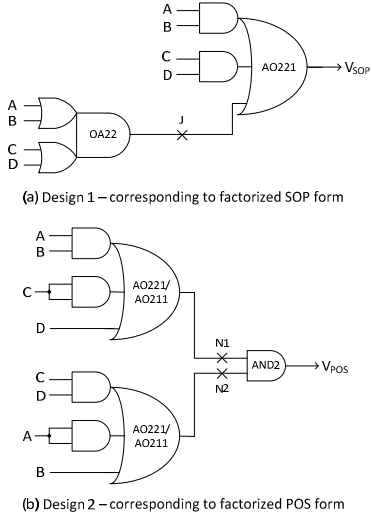

Figure 1. Synthesized combinational circuits: (a) based on factorized SOP form, and (b) based on factorized POS form.

A truth-cum-fault enumeration table was presented in [28] to estimate the intrinsic fault masking ratio (FMR) of a combinational circuit. Assuming that a combinational circuit has $n$ primary inputs and $m$ internal node(s), the total number of faulty and non-faulty combinations which have to be modelled would be specified by $2^{n+m}$. Of this, the non-faulty combinations would equate to $2^n$. It should be noted here that not all the internal fault(s) which may occur within a circuit may cause an output error [4] since some of the faults may be successfully masked by the circuit. Hence it is important to determine the ratio of the faults that will be masked, which will not cause any primary output(s) error, in proportion to the total number of faults that may potentially occur within the circuit. The FMR metric [28] precisely captures this phenomenon and gives a numerical estimate that can be used to quantify the intrinsic fault tolerance of a logic circuit, which may be combinational or sequential. It is to be noted that FMR is not used to account for faults that may occur at the primary inputs since any modification of the primary inputs basically implies a new input combination and the circuit is then expected to produce only that output that corresponds to the new input combination, and this is not considered to be a faulty operation of the circuit.

Now presuming that $k$ internal faulty combinations (not $k$ internal faults) which may occur within a circuit do not cause any primary output errors, we can obtain a generalized expression for FMR, given by (4). In (4), if $k$ equals 0, then FMR would also equate to 0, which implies that all the internal faults occurring within the circuit are said to be exposed since they cause primary output errors. Supposing if the numerator and denominator of (4) are equal, it implies that none of the internal circuit faults result in any primary output error and hence the circuit is said to be 100% robust. Hence, FMR = 1 represents the best-case, and the worst-case is FMR = 0, and practically, $0 \le \text{FMR} \le 1$. Hence, higher the FMR, better is the fault tolerance of a logic circuit.

$$\text{FMR} = \frac{k}{2^{n+m} - 2^n} \qquad (4)$$

TABLE II. TRUTH-CUM-FAULT ENUMERATION TABLE OF THE COMBINATIONAL CIRCUIT SHOWN IN FIGURE 1A

| Primary Inputs | | | | Internal Output | Primary Output (PO) | State of PO |
|---|---|---|---|---|---|---|
| A | B | C | D | J = (A+B) (C+D) | $V_{SOP}$ = AB+CD+J | |
| 0 | 0 | 0 | 0 | 0 | 0 | Actual |
| | | | | 1 (0→1 fault) | 1 | Error |
| 0 | 0 | 0 | 1 | 0 | 0 | Actual |
| | | | | 1 (0→1 fault) | 1 | Error |
| 0 | 0 | 1 | 0 | 0 | 0 | Actual |
| | | | | 1 (0→1 fault) | 1 | Error |
| 0 | 0 | 1 | 1 | 0 | 1 | Actual |
| | | | | 1 (0→1 fault) | 1 | Correct |
| 0 | 1 | 0 | 0 | 0 | 0 | Actual |
| | | | | 1 (0→1 fault) | 1 | Error |
| 0 | 1 | 0 | 1 | 1 | 1 | Actual |
| | | | | 0 (1→0 fault) | 0 | Error |
| 0 | 1 | 1 | 0 | 1 | 1 | Actual |
| | | | | 0 (1→0 fault) | 0 | Error |
| 0 | 1 | 1 | 1 | 1 | 1 | Actual |
| | | | | 0 (1→0 fault) | 1 | Correct |
| 1 | 0 | 0 | 0 | 0 | 0 | Actual |
| | | | | 1 (0→1 fault) | 1 | Error |
| 1 | 0 | 0 | 1 | 1 | 1 | Actual |
| | | | | 0 (1→0 fault) | 0 | Error |
| 1 | 0 | 1 | 0 | 1 | 1 | Actual |
| | | | | 0 (1→0 fault) | 0 | Error |
| 1 | 0 | 1 | 1 | 1 | 1 | Actual |
| | | | | 0 (1→0 fault) | 1 | Correct |
| 1 | 1 | 0 | 0 | 0 | 1 | Actual |
| | | | | 1 (0→1 fault) | 1 | Correct |
| 1 | 1 | 0 | 1 | 1 | 1 | Actual |
| | | | | 0 (1→0 fault) | 1 | Correct |
| 1 | 1 | 1 | 0 | 1 | 1 | Actual |
| | | | | 0 (1→0 fault) | 1 | Correct |
| 1 | 1 | 1 | 1 | 1 | 1 | Actual |
| | | | | 0 (1→0 fault) | 1 | Correct |

Table II shows the truth-cum-fault enumeration table of the combinational circuit shown in Figure 1a, based on [28]. Table II specifies all possible distinct input combinations corresponding to the primary inputs, the internal output, and the corresponding value of the primary output ($V_{SOP}$) for the non-faulty and faulty internal output scenarios. A fault may occur on the internal output J in Figure 1a in either direction viz. a 0→1 fault or a 1→0 fault which may occur either due to a transient (i.e. temporary) or a permanent effect [33]. Reference [18] had proposed the use of 0→1 and 1→0 symbols to model the transient or permanent (i.e. stuck-at) faults which may occur in logic circuits, and herein we adopt the same notations similar to that of [28]. The state of the primary output, whether it is actual (i.e. expected), correct, or an error is mentioned in the last column of Table II. The blackened boxes in Table II highlight the correct values of the internal circuit output which indicate no fault occurrence.

When the primary output (PO) state is correct it means that the internal circuit fault has been successfully masked, and when the PO state is an error, it means that the internal circuit fault is exposed. Hence, according to (4), FMR = 7/16 = 0.4375 for the combinational circuit shown in Figure 1a.

TABLE III. TRUTH-CUM-FAULT ENUMERATION TABLE OF THE COMBINATIONAL CIRCUIT SHOWN IN FIGURE 1B

| Primary Inputs | | | | Internal Outputs | | Primary Output (PO) | State of PO |
|---|---|---|---|---|---|---|---|
| A | B | C | D | N1 | N2 | $V_{POS}$ | |
| 0 | 0 | 0 | 0 | 0 | 0 | 0 | Actual |
| | | | | 0 | 1 (0→1) | 0 | Correct |
| | | | | 1 (0→1) | 0 | 0 | Correct |
| | | | | 1 (0→1) | 1 (0→1) | 1 | Error |
| 0 | 0 | 0 | 1 | 1 | 0 | 0 | Actual |
| | | | | 0 (1→0) | 0 | 0 | Correct |
| | | | | 0 (1→0) | 1 (0→1) | 0 | Correct |
| | | | | 1 | 1 (0→1) | 1 | Error |
| 0 | 0 | 1 | 0 | 1 | 0 | 0 | Actual |
| | | | | 0 (1→0) | 0 | 0 | Correct |
| | | | | 0 (1→0) | 1 (0→1) | 0 | Correct |
| | | | | 1 | 1 (0→1) | 1 | Error |
| 0 | 0 | 1 | 1 | 1 | 1 | 1 | Actual |
| | | | | 0 (1→0) | 0 (1→0) | 0 | Error |
| | | | | 0 (1→0) | 1 | 0 | Error |
| | | | | 1 | 0 (1→0) | 0 | Error |
| 0 | 1 | 0 | 0 | 0 | 1 | 0 | Actual |
| | | | | 0 | 0 (1→0) | 0 | Correct |
| | | | | 1 (0→1) | 0 (1→0) | 0 | Correct |
| | | | | 1 (0→1) | 1 | 1 | Error |
| 0 | 1 | 0 | 1 | 1 | 1 | 1 | Actual |
| | | | | 0 (1→0) | 0 (1→0) | 0 | Error |
| | | | | 0 (1→0) | 1 | 0 | Error |
| | | | | 1 | 0 (1→0) | 0 | Error |
| 0 | 1 | 1 | 0 | 1 | 1 | 1 | Actual |
| | | | | 0 (1→0) | 0 (1→0) | 0 | Error |
| | | | | 0 (1→0) | 1 | 0 | Error |
| | | | | 1 | 0 (1→0) | 0 | Error |
| 0 | 1 | 1 | 1 | 1 | 1 | 1 | Actual |
| | | | | 0 (1→0) | 0 (1→0) | 0 | Error |
| | | | | 0 (1→0) | 1 | 0 | Error |
| | | | | 1 | 0 (1→0) | 0 | Error |
| 1 | 0 | 0 | 0 | 0 | 1 | 0 | Actual |
| | | | | 0 | 0 (1→0) | 0 | Correct |
| | | | | 1 (0→1) | 0 (1→0) | 0 | Correct |
| | | | | 1 (0→1) | 1 | 1 | Error |
| 1 | 0 | 0 | 1 | 1 | 1 | 1 | Actual |
| | | | | 0 (1→0) | 0 (1→0) | 0 | Error |
| | | | | 0 (1→0) | 1 | 0 | Error |
| | | | | 1 | 0 (1→0) | 0 | Error |
| 1 | 0 | 1 | 0 | 1 | 1 | 1 | Actual |
| | | | | 0 (1→0) | 0 (1→0) | 0 | Error |
| | | | | 0 (1→0) | 1 | 0 | Error |
| | | | | 1 | 0 (1→0) | 0 | Error |
| 1 | 0 | 1 | 1 | 1 | 1 | 1 | Actual |
| | | | | 0 (1→0) | 0 (1→0) | 0 | Error |
| | | | | 0 (1→0) | 1 | 0 | Error |
| | | | | 1 | 0 (1→0) | 0 | Error |
| 1 | 1 | 0 | 0 | 1 | 1 | 1 | Actual |
| | | | | 0 (1→0) | 0 (1→0) | 0 | Error |
| | | | | 0 (1→0) | 1 | 0 | Error |
| | | | | 1 | 0 (1→0) | 0 | Error |
| 1 | 1 | 0 | 1 | 1 | 1 | 1 | Actual |
| | | | | 0 (1→0) | 0 (1→0) | 0 | Error |
| | | | | 0 (1→0) | 1 | 0 | Error |
| | | | | 1 | 0 (1→0) | 0 | Error |
| 1 | 1 | 1 | 0 | 1 | 1 | 1 | Actual |
| | | | | 0 (1→0) | 0 (1→0) | 0 | Error |
| | | | | 0 (1→0) | 1 | 0 | Error |
| | | | | 1 | 0 (1→0) | 0 | Error |
| 1 | 1 | 1 | 1 | 1 | 1 | 1 | Actual |
| | | | | 0 (1→0) | 0 (1→0) | 0 | Error |
| | | | | 0 (1→0) | 1 | 0 | Error |
| | | | | 1 | 0 (1→0) | 0 | Error |

Table III shows the truth-cum-fault enumeration table of the combinational circuit shown in Figure 1b. In Figure 1b, there are two internal outputs viz. N1 and N2, and hence fault modeling has to be considered with respect to these two intermediate outputs as shown in Table III. The blackened boxes in Table III represent the correct values of the internal circuit outputs for the specified primary inputs, and hence the non-blackened boxes represent the faulty internal output(s). Thus, according to (4), FMR = 10/48 = 0.2083 for the combinational circuit shown in Figure 1b.

Comparing the FMRs of the combinational circuits shown in Figures 1a and 1b, calculated based on Tables II and III, we find that the FMR of Figure 1a is greater than the FMR of Figure 1b by 110%. Since Figure 1a embeds greater intrinsic fault tolerance compared to Figure 1b, henceforth, we shall primarily consider the combinational circuit shown in Figure 1a and try to identify the redundant logic that may be introduced to enhance its intrinsic fault tolerance i.e. FMR without affecting its native functionality.

III. REDUNDANT LOGIC IDENTIFICATION AND INSERTION

Referring to Table II, we see that there are scenarios where the internal output fault does not cause any primary output error due to inherent fault masking, and there are scenarios where the internal output fault causes a primary output error. Two types of primary output errors can be noticed in Table II: i) the primary output becomes 1 when it should be 0, and ii) the primary output becomes 0 when it should be 1. In this work, we focus on tackling the latter issue, i.e. the identification and introduction of redundant logic to ensure that the primary circuit output does not incorrectly produce 0 when it is expected to produce 1. In fact, trying to devise a solution to overcome both the error conditions would be mutually exclusive since the circuit outputs would be complementary.

We now present Table II in a modified format to aid with the redundant logic identification that is labeled as Table IV. The error scenarios we aim to address with respect to the primary output V are highlighted using the blackened boxes in Table IV. The last column in Table IV mentions the 'redundant logic (i.e. product terms) identified', and the redundant product terms are also shown in blackened boxes. The term 'not applicable' is used for those scenarios where the primary output becomes 1 when it should be 0. The term 'not necessary' is used for those scenarios where the internal fault is successfully masked by the circuit due to its intrinsic fault tolerance.

In Table IV, we see that four redundant product terms are identified, namely BD, BC, AD, and AC. When these four product terms are added to (1), we find that the logic expression for V does not change due to the Boolean law of idempotency. Hence, when these four product terms are coupled with the AO22 gate functionality, it would result in the formation of a hypothetical complex gate 1, which is shown in dotted lines in Figure 2a. We label a gate as hypothetical when such a gate does not form a part of commercial digital standard cell libraries. If the circuit portrayed by Figure 2a is implemented in a full-custom design fashion, then all the (1→0) primary output errors would be eliminated, but the drawback with this approach is that the entire circuit design now has to be full-custom rather than being semi-custom which would involve a significant design effort and this would not be suitable for the design of even medium size combinational circuits. Notwithstanding, an observation made with respect to Figure 2a is that the FMR of the circuit increases to 0.6875, which implies an improvement of 57.1% in comparison with the FMR of the circuit shown in Figure 1a.

TABLE IV. MODIFIED TRUTH-CUM-FAULT ENUMERATION TABLE OF THE COMBINATIONAL CIRCUIT SHOWN IN FIGURE 1A

| Primary Inputs | | | | Internal Output | Primary Output (PO) | State of PO | Redundant Logic Identified (Products) |
|---|---|---|---|---|---|---|---|
| A | B | C | D | J | V | | |
| 0 | 0 | 0 | 0 | 0 | 0 | Actual | Not applicable |
| | | | | 1 (0→1 fault) | 1 | Error | |
| 0 | 0 | 0 | 1 | 0 | 0 | Actual | Not applicable |
| | | | | 1 (0→1 fault) | 1 | Error | |
| 0 | 0 | 1 | 0 | 0 | 0 | Actual | Not applicable |
| | | | | 1 (0→1 fault) | 1 | Error | |
| 0 | 0 | 1 | 1 | 0 | 1 | Actual | Not necessary |
| | | | | 1 (0→1 fault) | 1 | Correct | |
| 0 | 1 | 0 | 0 | 0 | 0 | Actual | Not applicable |
| | | | | 1 (0→1 fault) | 1 | Error | |
| 0 | 1 | 0 | 1 | 1 | 1 | Actual | BD |
| | | | | 0 (1→0 fault) | 0 | Error | |
| 0 | 1 | 1 | 0 | 1 | 1 | Actual | BC |
| | | | | 0 (1→0 fault) | 0 | Error | |
| 0 | 1 | 1 | 1 | 1 | 1 | Actual | Not necessary |
| | | | | 0 (1→0 fault) | 1 | Correct | |
| 1 | 0 | 0 | 0 | 0 | 0 | Actual | Not applicable |
| | | | | 1 (0→1 fault) | 1 | Error | |
| 1 | 0 | 0 | 1 | 1 | 1 | Actual | AD |
| | | | | 0 (1→0 fault) | 0 | Error | |
| 1 | 0 | 1 | 0 | 1 | 1 | Actual | AC |
| | | | | 0 (1→0 fault) | 0 | Error | |
| 1 | 0 | 1 | 1 | 1 | 1 | Actual | Not necessary |
| | | | | 0 (1→0 fault) | 1 | Correct | |
| 1 | 1 | 0 | 0 | 0 | 1 | Actual | Not necessary |
| | | | | 1 (0→1 fault) | 1 | Correct | |
| 1 | 1 | 0 | 1 | 1 | 1 | Actual | Not necessary |
| | | | | 0 (1→0 fault) | 1 | Correct | |
| 1 | 1 | 1 | 0 | 1 | 1 | Actual | Not necessary |
| | | | | 0 (1→0 fault) | 1 | Correct | |
| 1 | 1 | 1 | 1 | 1 | 1 | Actual | Not necessary |
| | | | | 0 (1→0 fault) | 1 | Correct | |

The other important observation is that Figure 2a may be replaced by Figure 2b since the hypothetical complex gate 2 of Figure 2b obviates the need for the OA22 gate. The hypothetical complex gate 2, theoretically, would have the ideal FMR value of 1. But beware practical combinational circuits generally consist of multiple inputs and outputs and the entire circuit cannot be designed as just one full-custom complex gate. An alternate, practically viable, and a possibly efficient approach would be to modify a combinational circuit by identifying and introducing logic redundancy into it in a semi-custom design fashion, as discussed in this paper. This approach would in fact make such a design methodology for redundant logic insertion in combinational circuits generic since the semi-custom design method is widely adopted for digital circuits and systems synthesis.

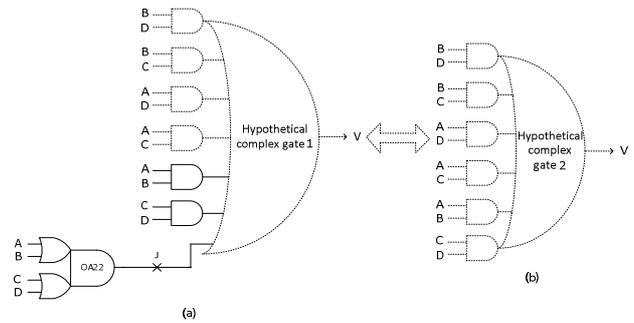

Figure 2. Alternative realizations of the combinational circuit case study using (a) a real complex gate and a hypothetical complex gate, and (b) one hypothetical complex gate.

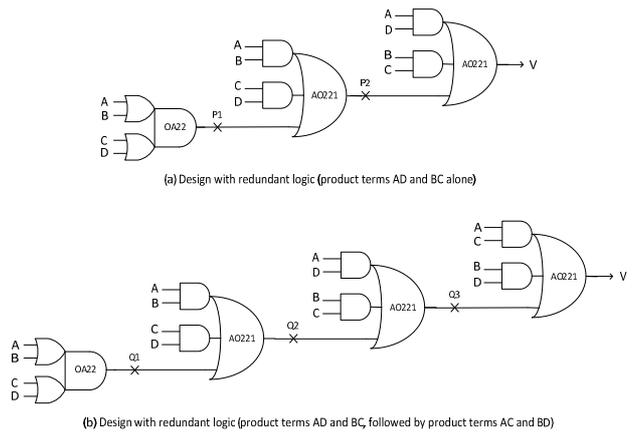

Figure 3. Combinational circuits with redundant logic.

However, due to the physical constraints of the digital cell library [31], not all the redundant product terms which are identified can be added simultaneously. So we gradually insert the redundant logic into the circuit shown in Figure 1a. For example, two redundant product terms, say AD and BC are inserted using an extra gate which is shown in Figure 3a, and the remaining redundant product terms viz. AC and BD are inserted using another gate as shown in Figure 3b. The choice of the redundant products insertion may not have a bearing on the fault masking capability though. However, note that the insertion of redundant logic would lead to an increase in the number of logic level(s) and internal outputs (nodes) compared to the original circuit shown in Figure 1a, and thus fault modeling has to be performed again for the entire circuit. Hence, the truth-cum-fault enumeration tables for the circuits shown in Figures 3a and 3b have been derived, and the new FMRs are also calculated. According to (4), for Figure 3a, FMR = 0.625, and for Figure 3b, FMR =

0.6786. Thus the FMRs of the circuits with redundant logic shown in Figures 3a and 3b are greater than the FMR of the circuit with no redundant logic that is shown in Figure 1a by 42.9% and 55.1% respectively. Thus it is clear that the identification and insertion of redundant logic into a combinational circuit helps to improve its fault tolerance. Nevertheless, this would be at the expense of some trade-off in the design metrics viz. delay, area, and power due to the addition of extra gate(s) to the original irredundant circuit.

Alternately, we intend to examine a hypothesis whether a circuit implementation that features logic sharing among its constituent terms would facilitate greater fault tolerance than a circuit implementation that synthesizes the factorized form and what happens if redundant logic is added on top of that. Here, we intend to analyze whether the direct physical implementation of the SOP expression of the combinational circuit case study would lead to an improved fault tolerance than the logically equivalent physical realization of the factorized SOP expression. We consider this because logic sharing may help to improve the intrinsic fault masking capability of a combinational circuit especially when the shared logic is preserved through physical implementation.

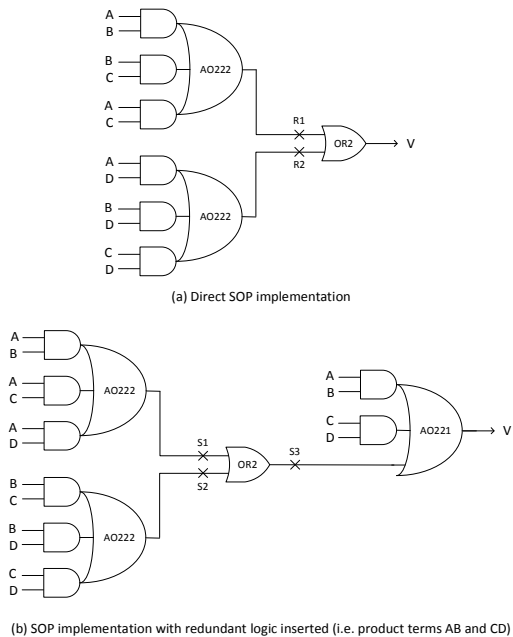

Figure 4. Additional redundant implementations of the combinational circuit.

The SOP form of the combinational circuit considered for our case study, given by (1), contains logic sharing. For example the minterm ABCD is covered by all the six product terms given in (1), and the physical synthesis of (1) is shown in Figure 4a. Considering the first AO222 gate shown in Figure 4a, the product terms AB, BC and AC are logically summed up using a single complex gate, and so the logic sharing among these product terms would be preserved. Alternatively, if the product terms AB, BC and AC are implemented using three 2-input AND gates and then if their outputs are combined using a 3-input OR gate, then three internal outputs would arise and when these are subject to fault(s), the logic sharing existent between the product terms may no more be realized and consequently this may not lead to an improvement in the fault tolerance. Figure 4b is an equivalent logic implementation of Figure 4a, but with explicit redundant logic introduced in the form of product terms AB and CD through the final AO221 gate. It is expected that the circuit shown in Figure 4b would exhibit greater fault tolerance than the circuit shown in Figure 4a due to redundant logic insertion.

Figure 4a shows a direct implementation of the SOP form of the combinational circuit given by (1), with logic sharing between the product terms AB, BC and AC preserved through the first AO222 gate, and the logic sharing between the remaining product terms AD, BD and CD preserved through the second AO222 gate. Note that there is no complex gate in the cell library that would realize the entire SOP form (1) as such.

The truth-cum-fault enumeration table for Figures 4a was derived, and based on that the redundant logic was identified as described earlier through Table IV. Figure 4b is an extended version of Figure 4a with partial redundant logic (i.e. product terms AB and CD) introduced. Subsequently, as per (4), the FMRs of the combinational circuits shown in Figures 4a and 4b were evaluated. The FMR of the combinational circuit shown in Figure 4a is estimated to be 0.4583, which is 4.8% greater than the FMR of the combinational circuit shown in Figure 1a, which validates our hypothesis. Nevertheless, this may be at the expense of some penalty in the design metrics since Figure 4a has an extra gate compared to Figure 1a. On the other hand, the FMR of the circuit shown in Figure 4b was estimated to be 0.6786, which is 48.1% greater than the FMR of the circuit shown in Figure 4a. This once again confirms that redundant logic insertion leads to greater fault tolerance, albeit involving some potential trade-off with respect to the design metrics. The FMRs of the combinational circuits shown in Figures 3b and 4b are identical (i.e. 0.6786). However, the logic depth of the latter is less than the logic depth of the former by one. Hence it is expected that Figure 4b when physically realized would have reduced propagation delay compared to the physical realization of Figure 3b, and this is substantiated by the simulation results presented in the next section.

IV. DESIGN METRICS AND FMRs OF VARIOUS CIRCUIT IMPLEMENTATIONS

The combinational circuits shown in Figures 1, 3 and 4 were technology mapped to the elements of the 32/28nm CMOS cell library [31]. They are referred by their respective figure numbers in Table V. Random input sequences were supplied to the different circuits to verify their functionalities and also to capture their switching activities. The switching activity files generated were subsequently used for average power estimation using Synopsys tools. Also, the maximum propagation delay and cells area occupancy were estimated. The design metrics are given in Table V, and the FMR of the different circuits evaluated are also given in the table.

To comprehensively comment on the design metrics, a figure-of-merit (FoM) is used, which is calculated as the inverse product of power and delay. The power-delay product (PDP) is a standard metric [34] used for evaluating digital circuit and system designs, and a low PDP value is desirable. Thus, a high value of FoM indicates an optimized design since FoM = $PDP^{-1}$.

From Table V, we see that the FoM of Figure 1a is the highest since it has the least number of gates compared to the other implementations. The FoM of Figure 1b is less than the FoM of Figure 1a by about 23%. Figures 3a, 3b, 4a and 4b contain extra gates i.e. redundant logic and so their FoMs would obviously be low compared to the FoM of Figure 1a. However, the FMRs of Figures 3b and 4b are the highest compared to the rest. Also, it is seen that the FoM of Figure 4b is better than the FoM of Figure 3b by 21.5%, while both feature the same FMR. The logic depth of Figure 4b is 3, and the logic depth of Figure 3b is 4, and so the propagation delay of the latter is expected to be more than the former. Figure 4b has 3 complex gates and 1 simple gate whereas Figure 3b has 4 complex gates. Although their areas are the same, nevertheless their power dissipations vary with Figure 4b dissipating slightly lesser power than Figure 3b. Due to these, the FoM of Figure 4b is more optimized compared to the FoM of Figure 3b. The circuit shown in Figure 1a is preferable from the FoM perspective, and the circuit shown in Figure 4b is preferable from the FMR perspective. Hence a clear trade-off between FoM and FMR is evident.

TABLE V. DESIGN METRICS AND FMRS CORRESPONDING TO DIFFERENT IMPLEMENTATIONS OF THE COMBINATIONAL CIRCUIT CASE STUDY

| Circuit Reference | Power (µW) | Delay (ns) | Area (µm$^2$) | FoM | FMR |
|---|---|---|---|---|---|
| Figure 1a | 1.681 | 0.15 | 5.59 | 3.9651 | 0.4375 |
| Figure 1b | 2.044 | 0.16 | 8.13 | 3.0581 | 0.2083 |
| Figure 3a | 2.569 | 0.23 | 8.64 | 1.6923 | 0.625 |
| Figure 3b | 3.408 | 0.31 | 11.69 | 0.9465 | 0.6786 |
| Figure 4a | 3.866 | 0.19 | 8.64 | 1.3615 | 0.4583 |
| Figure 4b | 3.344 | 0.26 | 11.69 | 1.1502 | 0.6786 |

## V. CONCLUSION AND SCOPE FOR FURTHER WORK

A novel redundant logic insertion scheme to improve the intrinsic fault tolerance of combinational circuits was presented in this work by taking into account the physical constraints of a digital cell library. Instead of replicating the entire circuit as such, the proposed scheme identifies the redundant logic that may introduced into the original circuit to improve its intrinsic fault tolerance as desired. Example circuit implementations which do not have and which have logic sharing preserved in them were considered for the analysis. The design parameters estimated and the FMRs calculated show that a trade-off is inevitable with respect to FoM and FMR. However, a possibly good choice for nanoelectronic designs might be to select an implementation that achieves a good compromise between FoM and FMR.

Since this paper has considered just a case study based analysis of the proposed scheme, there exists a significant scope for further work in terms of devising an optimized circuit synthesis algorithm to automate the electronic design that would be geared towards performing fault tolerance analysis efficiently and enabling circuit realizations which would achieve a good FMR and FoM tradeoff. But this may require coping up with a significant computational complexity with respect to the fault tolerance analysis. Presuming that a combinational circuit features $m$ internal outputs, the computational complexity associated with the fault modeling and analysis would be given by $O(2^m)$, which signifies an exponential complexity and might result in a state-space explosion if $m$ is large. To alleviate this problem, the entire circuit can be split up into as many sub-circuits as desired and the fault tolerances of the individual sub-circuits may be analyzed with reduced computational complexity. This method of partitioning a big circuit into small sub-circuits to perform the fault tolerance analysis may be less cumbersome and may significantly reduce the computational complexity and the simulation time involved. Nevertheless, care must be taken to ensure that the fault tolerance analysis of the sub-circuits can be subsequently combined correctly to account for the fault tolerance of the entire circuit. Hence this provides scope for a potential future work with respect to fault tolerance estimation and enhancement of fault tolerance of combinational and sequential circuits which may comprise an arbitrary number of inputs and outputs.